\documentclass{article}
\usepackage[utf8]{inputenc}
\usepackage{graphicx}
\usepackage{todonotes}
\usepackage{subcaption}
\usepackage{multirow}
\usepackage{hyperref}

\newcommand\blfootnote[1]{%
  \begingroup
  \renewcommand\thefootnote{}\footnote{#1}%
  \addtocounter{footnote}{-1}%
  \endgroup
}

\usepackage{multirow}
\usepackage{longtable}

\title{A Survey of Security Concerns and Countermeasures in Modern Micro-architectures with Transient Execution}
\author{Nikhilesh Singh, Vinod Ganesan, and Chester Rebeiro\\
Indian Institute of Technology Madras\\
\{\em nik,vinodg,chester\}@cse.iitm.ac.in}
\date{}

\begin{document}

\maketitle
\begin{abstract}

\textcolor{black}{
In the last two decades, the evolving cyber-threat landscape has brought to center stage the contentious tradeoffs between the security and performance of modern microprocessors. 
The guarantees provided by the hardware to ensure no violation of process boundaries have been shown to be breached in several real-world scenarios. While modern CPU features such as superscalar, out-of-order, simultaneous multi-threading, and speculative execution play a critical role in boosting system performance, they are central for a potent class of security attacks termed transient micro-architectural attacks.
These attacks leverage shared hardware resources in the CPU that are used during speculative and out-of-order execution to steal sensitive information. Researchers have used these attacks to read data from the Operating Systems (OS) and Trusted Execution Environments (TEE) and to even break hardware-enforced isolation.}

\textcolor{black}{Over the years, several variants of transient micro-architectural attacks have been developed. While each variant differs in the shared hardware resource used, the underlying attack follows a similar strategy. 
This paper presents a panoramic view of security concerns in modern CPUs, focusing on the mechanisms of these attacks and providing a classification of the variants. Further, we discuss state-of-the-art defense mechanisms towards mitigating these attacks. }
\end{abstract}

\blfootnote{This is the authors' version of the book chapter \textit{Secure Processor Architectures
 } published in the \textbf{Handbook of Computer Architecture}, \textit{Living Edition} by \textit{Springer}, available at \href{https://doi.org/10.1007/978-981-15-6401-7_10-1}{https://doi.org/10.1007/978-981-15-6401-7$\_$10-1} }

\section{Introduction}\label{sec:intro}
For over half a century, microprocessor research has focused on improving performance. Various micro-architectural features such as cache memories, branch prediction, superscalar, speculative, and out-of-order execution were developed to facilitate this. While some of these features, for example,  the cache memory, were introduced to hide the latency of slow components, others like branch predictors, helped hide overheads due to operations that  slow down program execution. Features like out-of-order execution and speculative execution were introduced to better utilize available resources. 
Side-by-side, features were incorporated in processors to support better multi-programming. Features such as multi-core processors and hardware multi-threading were incorporated to allow multiple users to simultaneously share a processor. These features accelerated new computing paradigms, especially cloud computing, where multiple users simultaneously share common hardware, thereby drastically reducing computation costs.

A critical aspect of the cloud computing paradigm is the isolation between users. To isolate one user's program from another, security schemes such as protection rings, segmentation, page table access control bits, virtualization support, hardware-based security, crypto-accelerators, and trusted execution environments were introduced. 
Very soon, it was realized that these security schemes were insufficient. The shared hardware became a source of information leaks that could undermine the isolation provided by the processor. 
These attacks, popularly known as {\em micro-architectural attacks}, made use of shared hardware resources to glean sensitive information such as cryptographic keys, web pages visited, user passwords, and keystrokes. Different strategies such as time-driven attacks, Prime+Probe, Flush+Reload, and Evict+Time were proposed for this purpose. In a cloud computing environment, these attacks could leak information from one user to another, in spite of having all security features enabled.  

In 2018, two potent micro-architectural attack variants were proposed, namely Meltdown~\cite{Lipp:2018:Meltdown} and Spectre~\cite{Kocher:2019:Spectre}, that exploited the speculative and out-of-order execution features present in microprocessors. These attacks leveraged the fact that a processor's speculation may not always be correct. When speculation goes wrong, the speculatively executed instructions, called transient instructions, need to be discarded, and the CPU should be rolled back to a previous state. However, this rollback is not always perfect. The CPU would still have a reminisce of the transient instructions. Researchers showed how this reminisce  can be used to leak secrets. These attacks, which came to be called {\em transient micro-architectural attacks}, could read the contents of any memory region, including the OS memory. It could also read memory from trusted  enclaves, even though the enclaves used encrypted memory.  

Since 2018,  there been several variants of transient micro-architectural attacks including Zombieload~\cite{Schwarz:2019:Zombieload}, Foreshadow~\cite{Bulck:2018:Foreshadow}, Rogue In-Flight Data Load (RIDL)~\cite{Schaik:2019:RIDL}, Fallout~\cite{Canella:2019:Fallout}, Load Value Injection (LVI)~\cite{Bulck:2020:LVI}, and Crosstalk~\cite{Ragab:2021:Crosstalk}. Each variant found a new vulnerability that could bypass isolation in the CPU. Many of these attacks are not easily prevented by software patches. For those that can, the patches have huge performance penalties. It would require fundamental changes in the CPU design to mitigate these attacks in hardware. 

This paper provides an introduction to transient micro-architectural attacks. Starting from Meltdown and Spectre, we dwell on the basic principle of the attacks. This would be useful in distinguishing between the various attack classes and discussing the available mitigation techniques. Section~\ref{sec:uarch} provides a background of modern CPU micro-architecture and also gives an introduction to micro-architectural attacks. Section~\ref{sec:tma} discusses  transient micro-architectural attacks and classifies them. Section~\ref{sec:defenses} discusses the defenses for these attacks, while the final section has the concluding remarks.

\section{Modern CPU Microarchitecture\label{sec:uarch}}

\begin{figure}[h!]
    \centering
    \includegraphics[width=0.9\linewidth]{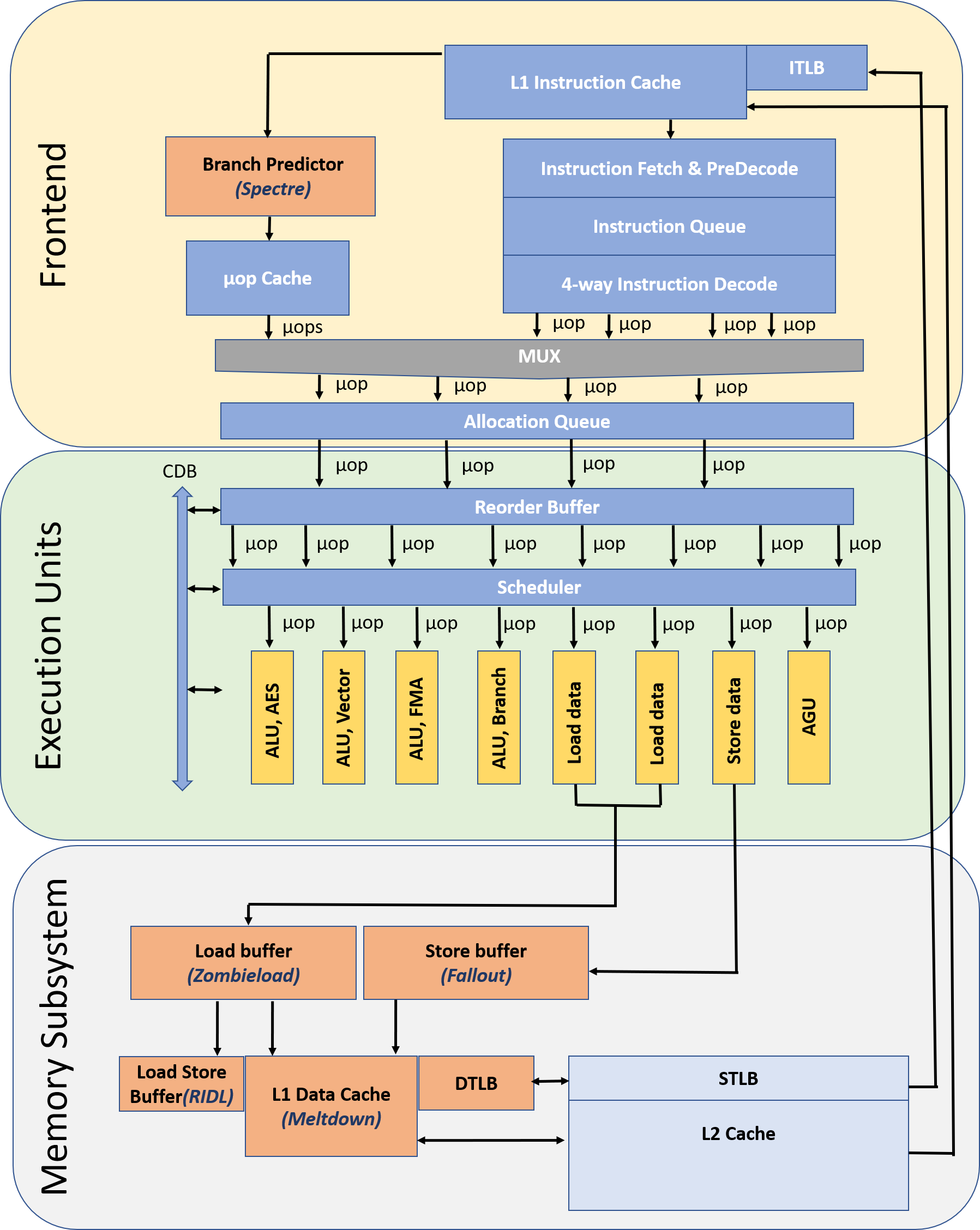}
    \caption{An Out-of-order Superscalar processor with vulnerable components shaded in Orange.}
    \label{fig:ooo}
\end{figure}

{\flushleft \bf Notions of Security in Microprocesosrs.}
\textcolor{black}{
Beyond functional correctness, modern microprocessors attempt to enforce a root of trust to mitigate the ever-growing array of attacks. The goal of such approaches is to enable secure booting and provide a platform to launch Trusted Execution Environments (TEE) post boot up. These TEEs, such as ARM Trustzone and Intel Software Guard Extensions (SGX)~\cite{Schunter:2016:IntelSGX}, ensure that the process boundaries guaranteed by the hardware are not violated by other processes.
For example, the Intel SGX adopted in 2015, is a TEE feature supported by commercial processors 
that provide private regions of memory for programs. These regions are known as enclaves, and cannot be accessed even from privileged software like the Operating System. This is achieved by encrypting enclave code and data present in the DRAM. Decryption is done when the code or data is fetched into the processor. Thus, the contents of an enclave, when in RAM, are always in an encrypted form and not accessible to any code outside the enclave regardless of the privilege levels. In recent years, however, researchers have shown that such trusted execution environments are not a panacea against the threat of transient micro-architectural attacks~\cite{Bulck:2018:Foreshadow,Weisse:2018:ForeshadowNG}. The potency of these attacks is one of the reasons that led to the deprecation of Intel SGX from upcoming desktop processors~\cite{Intel:2021:Core11Datasheet,Intel:2022:Core12Datasheet}, posing further open questions regarding the security of hardware designs. In this section, we explore the premise of such attacks on the micro-architecture from first principles, starting with a background on the working of transient instruction in superscalar CPUs.}

{\flushleft \bf Transient Instructions in Superscalar CPUs.} Figure~\ref{fig:ooo} shows
a block diagram of a superscalar CPU.
In every clock cycle, multiple instructions are fetched from the
instruction cache into an instruction/decode buffer which forms the frontend. The instructions are decoded 
into a set of micro-ops and are continuously fed to the exeuction engines, such as Arithmetic and Logic Units (ALU) and Floating Point Units (FPU), through a dispatcher of allocation queues. 
The scheduler ensures that the issue is possible only if the functional unit is available and the operands used by the instruction are up-to-date.

The instructions to the functional units can be issued out-of-order and based on a speculation, for example, the CPU can predict the outcome of a branch and speculatively execute instructions at the predicted branch target.  
 The results  from these speculatively executed instructions are stored in a temporary buffer and committed to registers and memory only when the speculation turns out correct. 
 On the other hand, if the speculation is wrong, for example, a branch is mis-predicted, the results from the speculatively executed instructions are dropped and not committed. These instructions are known as {\em transient instructions}. Besides branch mis-predictions there are several reasons that can cause transient instructions. For instance,  a user-space program executing a load or store instruction from an illegal memory, for example from the kernel space, can result in a memory exception and also transient instructions. Another instance is of bounds check instructions that identify if an index is within array bounds. Memory operations following the bounds can be speculatively executed with any arbitrary out-of-bound index. 
 
 In addition to the out-of-order and speculative execution of processes, many modern CPUs support the  execution of multiple programs simultaneously. This feature is known as Symmetrical Multi-threading (SMT). Instructions from two or more programs simultaneously execute in a single pipeline sharing hardware resources such as cache memories, branch prediction units, and various other on-chip resources. 

{\flushleft \bf Micro-architectural State.}
As instructions flow through the CPU, various registers, buffers, caches, and other memory structures in the CPU core store temporary data and results from the execution. While a few of these memory structures, for instance, the general purpose registers, can be read or modified using instructions by instructions in the ISA, a significant portion of the   structures are  hidden and inaccessible from   software. 
To enforce separation between applications, system software ensures that the data  present in the ISA visible shared memory structures of one application cannot be read or modified by another application. For example, during a context switch,  general purpose registers are either invalidated or loaded with the context of the next process that executes, thus achieving a temporal separation between  the two processes. In  multi-core or multi-threaded CPUs on the other hand, the ISA visible memory structures are duplicated enforcing spatial separation. 

Unlike the visible  structures, the hidden memory structures in the CPU, such as cache memories and branch prediction units are not always spatially and temporally separated between applications. 
They retain their values across context switches and are possibly  shared in multi-core and multi-threaded CPUs. 
For example, a cache line that holds data from one application, can be evicted by another application. Similarly, a branch predictor trained on branches in one application can influence the outcome of a prediction in another application. 
At first glance, this may seem innocuous as the structures are hidden from software. However, researchers have found that one application can indirectly affect another through these shared hidden memory structures. This has led to a series of security vulnerabilities, commonly grouped in a category called micro-architectural attacks. The red regions in Figure~\ref{fig:ooo} are modules in the processor with demonstrated security vulnerabilities. Researchers have used these vulnerabilities to break cryptographic algorithms, read Operating System data and break trusted execution environments.

Researchers have used these vulnerabilities to break cryptographic algorithms~\cite{Bernstein:2005:CacheAttack,percival:05}, 
design keyloggers~\cite{Ristenpart:2009:Heyyou}, fingerprint websites~\cite{Shusterman:2019}, break security features like Address Space Layout Randomization~\cite{Barresi:2015:CAIN,Gras:2017:ASLR,Hund:2013:againstASLR}, leak sensitive information from the operating system~\cite{Kocher:2019:Spectre,Lipp:2018:Meltdown} and trusted enclaves~\cite{Bulck:2018:Foreshadow,Weisse:2018:ForeshadowNG}.  They have been applied on a variety of devices ranging from mobile phones to cloud computing servers.
The next section provides a brief introduction to micro-architectural attacks.

\subsection{Micro-architectural Attacks}\label{sec:micro-arch-attacks}
This section introduces micro-architectural attacks using the  example of cache memories. The cache memory is a  high-speed memory placed between the CPU and RAM to cache recently used instructions and data. It can be simultaneously shared by multiple applications in a CPU core. Due to its small size, it can be the cause of contention when  applications compete for the same cache line. The authors explain the fundamental working of micro-architectural attacks by using three examples. The first uses a prime and probe algorithm on a shared cache memory, while the second is an algorithm called flush and reload, that uses shared library code. The third is an evict and time algorithm on the cache memory.

\begin{figure}[!t]
\begin{subfigure}{\textwidth}
  \centering
  \includegraphics[width=.9\linewidth]{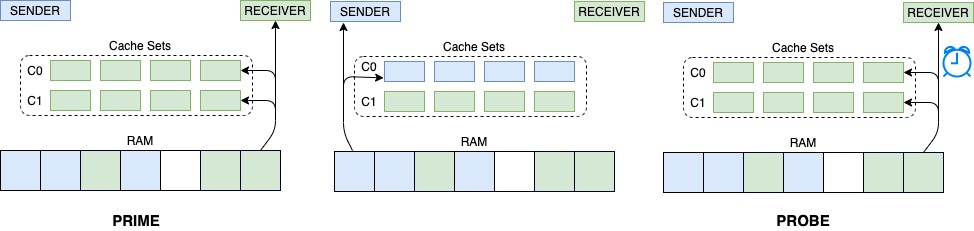}  
  \caption{Prime+Probe}
  \label{fig:pnp}
\end{subfigure}
\begin{subfigure}{\textwidth}
  \centering
  \includegraphics[width=.9\linewidth]{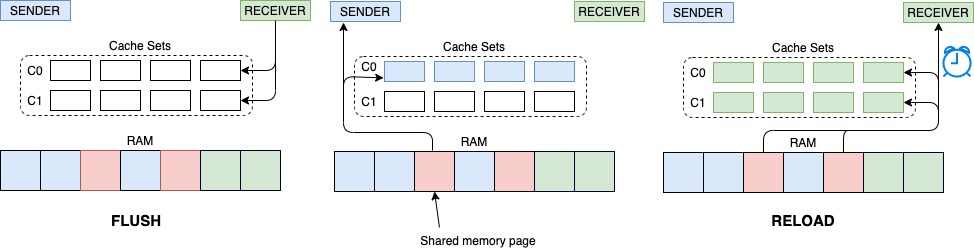}  
  \caption{Flush+Reload}
  \label{fig:fnr}
\end{subfigure}
\begin{subfigure}{\textwidth}
  \centering
  \includegraphics[width=.9\linewidth]{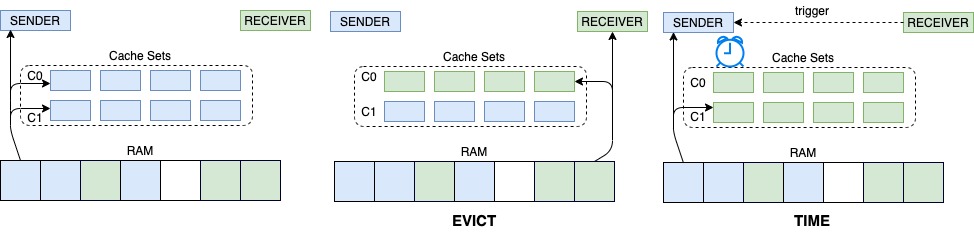}  
  \caption{Evict+Time}
  \label{fig:ent}
\end{subfigure}
\caption{Prime+Probe, Flush+Reload, and Evict+Time are the most common algorithms used to exfiltrate data in a micro-architectural attack. This figure demonstrates these algorithms in a covert channel that uses cache memory to transmit one bit of information from a sender to a receiver.}
\label{fig:uarch}
\end{figure}

{\flushleft \bf Prime+Probe Attacks.}
Prime and probe forms the basis for several micro-architectural attacks. It exploits the variance in the execution time caused by two applications that contend for the same shared hardware resource. The attack is discussed by showing an example of how the  cache memory can be used to create a covert communication channel between a high-privileged application and a low-privileged application. Similar channels have been used on other shared resources as well, such as TLBs, branch prediction units, load and store buffers, and even DRAM.

Consider that the high-privileged application  called the {\em sender} and the low-privileged application called the {\em receiver} share a common  cache memory. For example, the sender and receiver execute simultaneously on a CPU with a shared L1 data cache memory. The objective of the covert channel is to use the increase in execution time due to contention in the cache memory to transmit a message from the sender to the receiver. Apriori, the sender and receiver agree upon two cache sets $C0$ and $C1$ for communicating bits 0 and  1, respectively. The communication works as follows. {\bf (1)} The receiver first performs memory load operations that fill both cache sets. This is called the {\em prime} phase and is done by loading data from addresses that map to sets $C0$ and $C1$ as shown in Figure~\ref{fig:pnp}. {\bf (2)} Depending on the message bit, the sender performs a memory operation  to evict the receiver's data from the corresponding cache set. For example, to transmit a 0, the sender would evict the receiver's data from the cache set $C0$. {\bf (3)} In the {\em probe} phase, the receiver repeats the  memory operations in step (1), but this time also measures the execution time.  Based on the execution time, the receiver can infer the transmitted bit since the memory access to the evicted cache set would take longer owing to the cache miss.

Prime+probe in micro-architectural attacks work similarly, except for the fact that the sender and receiver do not collude. Instead, the receiver primes a sufficient number of sets in the cache (step (1)), waits for the sender to execute and evict one or more cache lines in these sets, and then performs a probe similar to step (3) to identify patterns in the sender's execution.

{\flushleft \bf Flush+Reload Attacks.}
Unlike Prime+Probe attacks, where the information leakage is due to conflicts in the cache memory, in the Flush+Reload attacks, information leakage is  caused without forcing cache conflicts. Consider, for instance, the high and low-privileged applications sharing  memory pages. Such sharing is common in systems that use shared libraries. A single copy of the shared library present in RAM is used by multiple applications. The time required to load data in a shared memory page depends on whether the data is in the cache or not. If present in the cache, the load will be considerably faster than if it is not present in the cache. 

Consider  that the sender and receiver of a covert channel decide on two  shared regions of code or data, for example in a shared library. These regions are chosen so that they map to distinct cache sets: $C0$ and $C1$. 
In step {\bf (1)}, the receiver  ensures that the data in these two regions are not in the cache shown in Figure~\ref{fig:fnr}. This is performed by a flush operation that evicts the addresses from the cache and is called the {\em flush} phase. On Intel x86 platforms, an instruction called {\tt clflush} is used to perform this. The {\tt clflush} takes an address as an argument and flushes the address from all caches in the CPU.
{\bf (2)} In the second step, the sender, performs load operations to either $C0$ or $C1$ depending on whether it wants to transmit a 0 or 1 respectively. It would cause the data from one of the two shared regions to be fetched into the cache. 
{\bf (3)} The receiver then performs loads on both addresses and measures the time taken. This is called the {\em reload} phase. Only one of these two loads would result in a cache hit. The time when there is a cache hit would be much shorter than the time when there is a cache miss. This difference in time can be used to infer the bit transmitted.
Unlike the Prime+Probe attacks techniques, Flush+Reload is independent of the cache attributes, like its associativity. It thus results in more portable attacks.

In transient micro-architectural attacks, the attacker defines an array. Similar to the covert channel, in step (1) the attacker ensures that no element of the array is present in the cache memory. In step (2), the attacker triggers a transient load operation that forces exactly one element from the array to be loaded into the cache. Similar to step (3) in the covert channel, the attacker would do a reload to identify which element was loaded. An element of the array that is loaded transiently often reveal secret information,  the Operating System data.

{\flushleft \bf Evict+Time Attacks.} Evict+Time attacks closely resemble the Prime+Probe attacks. The difference is that the adversary is able to accurately measure the execution time of the sender application. While this is a strong assumption, there are certain scenarios where such measurements are possible. For example, when the adversary can trigger the execution of the sender  and an observable event marks the end of its execution. In such cases, the duration between the trigger and the event serves as a measure of the execution time of the sender.

Consider again the covert channel between the high-privileged sender and low-privileged receiver application. The assumption at the start is that both cache sets, $C0$ and $C1$,  have the sender's data. In the second step, the receiver evicts one of the cache sets, say $C0$ as shown in Figure~\ref{fig:ent}. In the third step, it triggers the sender to execute and monitors the execution time of the sender. The sender would transmit a $0$ or $1$ by loading data from memory that maps to the $C0$ and $C1$ cache set respectively. The time taken to perform this load differs for the $0$ and $1$ bit transmissions. transmitting $0$ will result in a cache miss, thus experiencing a longer execution time compared to transmitting $1$. This difference in time is observed by the receiver to infer the transmitted bit.

\section{Transient Micro-architectural Attacks}\label{sec:tma}
\begin{figure}[!t]
    \centering
    \includegraphics[width=0.95\linewidth]{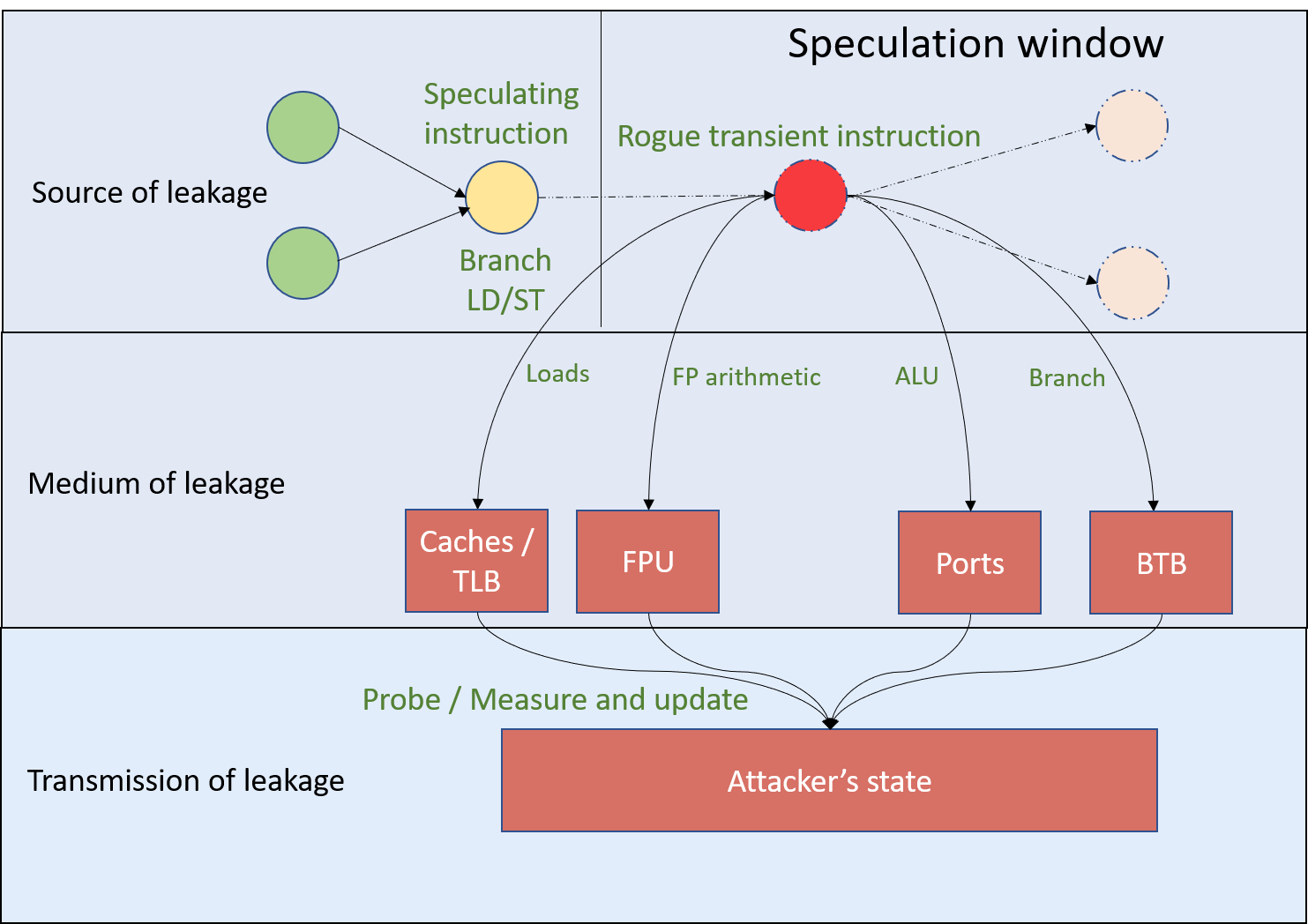}
    \caption{In a transient attack, the transient instruction modifies the hidden states of the CPU like cache memories, FPU, and ports,  in a manner that depends on secret information. In the next stage, the attacker exfiltrates these secrets from the hidden states.}
    \label{fig:transient}
\end{figure}

When transient instructions execute, the hidden state of the CPU is modified. While the results of a transient operation are discarded after the speculation is proved wrong, the hidden state of the CPU is not rolled back. Thus, transient instructions have a permanent impact on the CPU state. Consider, for example, the following code snippet.
\begin{center}
\begin{verbatim}
    I1. cmp r0, r1    
    I2. jne <dest_addr> /* branch to dest_addr, if r0 != r1 */
    I3. mov r2, Addr1 
    I4. add r2, r1      /* r2  = r2 + r1 */
    I5. load r3, r2     /* r3  = memory corresponding to (r2) */ 
\end{verbatim}
\end{center}
In an out-of-order processor,  instructions {\tt I3}, {\tt I4}, and {\tt I5} can be transiently executed if the CPU mispredicts the branch outcome at {\tt I2}. If the memory load in {\tt I5} results in a cache miss, it causes data at the address present in {\tt r2} to be loaded into cache. Due to the misprediction, the CPU would discard the results of instruction {\tt I3}, {\tt I4}, and {\tt I5}; however, it would not roll back the state of the cache memory. Thus, data corresponding to the memory load in {\tt I5} would persist in the cache even after the transient executions are dropped. In 2018, researchers showed that this reminiscence of a transient execution could lead to serious security vulnerabilities that could potentially compromise every application executing on the CPU. The two attacks, namely Meltdown and Spectre, that were proposed in 2018 showed how this reminiscence could undermine the security of application software on a variety of commercial microprocessors. Since then, several variants of such transient attacks have been proposed. They form a new class of extremely powerful micro-architectural attacks and have come to be known as {transient micro-architectural attacks} or simply {\em transient attacks}.

A typical transient attack has three stages, as shown in Figure~\ref{fig:transient}. The first stage disrupts the flow of program execution by forcing an exception or by inducing a misprediction that could trigger transient execution. In the next stage, the attacker relies on one or more of the transiently executed instructions to modify a hidden CPU state, such as the cache memory, branch predictor, or an internal buffer. The transient instruction is designed in a way so that the modification in the hidden state is correlated with a secret. The secret, for instance, can be keys of cryptographic ciphers, kernel code or data regions, or any other sensitive information. Due to the exception or the misprediction that occurred in the second phase, the transiently executed instructions are discarded, while the hidden micro-architectural states remain unaltered. In the final stage, the attacker exfiltrates information from the hidden micro-architectural state using an algorithm like Prime+Probe, Evict+Time, or Flush+Reload, to glean information about the secret.

\begin{table*}[!t]
\small
\begin{center}
\caption{Transitive Micro-architectural Attacks\label{tab:tma}}
  \centering
    \begin{tabular}{|p{.16\textwidth}|c|l|}
    \hline
    {\bf Attack} & {\bf Requirement} &  {\bf Source of Leakage} \\
    \hline
    \multicolumn{3}{|c|}{|Meltdown and Spectre Like Attacks|}\\
    \hline
    Meltdown~\cite{Lipp:2018:Meltdown}    & \multirow{3}{*}{SMT} & Transitive load \\ \cline{1-1}\cline{3-3}
    Spectre~\cite{Kocher:2019:Spectre}     &                      & BPU \\ \cline{1-1}\cline{3-3}
    Foreshadow~\cite{Bulck:2018:Foreshadow}  &                      & Transitive load \\
    \hline
    \multicolumn{3}{|c|}{|Micro-architectural Data Sampling|}\\
    \hline
    RIDL~\cite{Schaik:2019:RIDL}        & \multirow{3}{*}{SMT} & Line feed buffer \\ \cline{1-1}\cline{3-3}
    Fallout~\cite{Canella:2019:Fallout}     &                      & Store buffer  \\ \cline{1-1}\cline{3-3}
    Zombieload~\cite{Schwarz:2019:Zombieload}  &                      & Line feed buffer \\ \cline{1-1}\cline{3-3}
    LVI~\cite{Bulck:2020:LVI}         &                      & {Store buffer} \\ \cline{1-1}\cline{3-3}
    Crosstalk~\cite{Ragab:2021:Crosstalk}   &                      & Staging buffer \\
    \hline
\end{tabular}
\end{center}
\end{table*}

After the initial attacks, vis-\`{a}-vis Meltdown and Spectre, several variants of transient attacks have appeared in the literature~\cite{Bhattacharyya:2019:SMoTherSpectre,Bulck:2018:Foreshadow,Bulck:2020:LVI,Canella:2019:Fallout, Chen:2020:SGXPectre,Ragab:2021:Crosstalk,Schwarz:2019:Zombieload, Schwartz:2019:NetSpectre,Schaik:2019:RIDL,Weisse:2018:ForeshadowNG}. Each new attack identified a new medium of leakage. Broadly, these attacks can be categorized into two classes based on the micro-architectural medium used for the leakage. The first is address-controllable transient attacks like Meltdown and Spectre, 
while the others are based on micro-architectural data sampling from internal buffers.
 While at a high level, the stages in both categories are the same and follow Figure~\ref{fig:transient}, there are subtle differences between the two classes. Address-dependent attacks like Meltdown and Spectre use micro-architectural components like cache memories or branch prediction units as a medium for leakage. In these attacks, data (or instructions) placed in strategic memory addresses are transiently loaded (or executed). For example, in the covert channels described in Section~\ref{sec:micro-arch-attacks}, an address is used to select a cache set. The choice of the cache set is used as a medium for information leakage.  In micro-architectural data sampling attacks like Zombieload and Crosstalk, on the other hand, it is not the address that is critical. Instructions are crafted so as to snoop into internal buffers like Re-order buffers, Line-Fill buffers,  and load and store buffers. Table~\ref{tab:tma} classifies the known attacks into these two categories. 

\subsection{Meltdown and Spectre like Attacks}
These attacks require the knowledge of memory regions of interest, and the attacker can target them specifically. Attacks like Meltdown~\cite{Lipp:2018:Meltdown}, Spectre~\cite{Kocher:2019:Spectre} and Foreshadow~\cite{Bulck:2018:Foreshadow} fall in to this category. The upcoming sections look into each of these attacks to elaborate on their design and mechanisms.

{\flushleft \bf Meltdown.} CPUs use protection rings to isolate privileged code. For example, Intel CPUs have four rings: Ring 0 to Ring 3. Privileged code, such as the Operating System's kernel, is assigned to Ring 0, while user processes are assigned to Ring 3. The hardware ensures that during regular operations, code executing in Ring 3 cannot read or write to Ring 0, thus isolating the kernel's code and data from userspace programs.
The Meltdown attack exploits transient execution to read kernel data from a user program, thus breaching the isolation provided by the protection rings. 

\begin{figure}
    \centering
    \includegraphics[width=10cm]{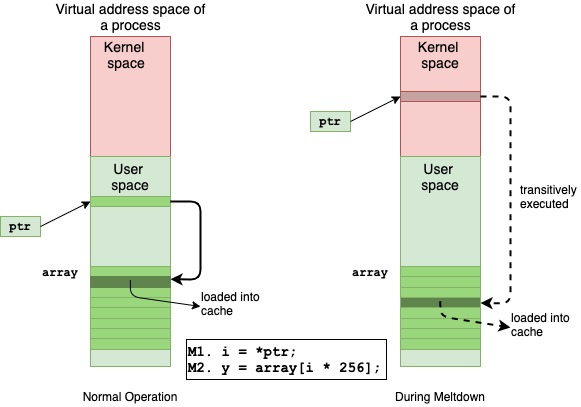}
    \caption{Transitive execution of a memory load instruction causes data from {\tt array} to be loaded into the cache memory. Unlike the visible micro-architectural state, the cache contents are not rolled back when transient instructions are discarded. The contents of the cache can be gleaned using techniques such as Prime+Probe or Flush+Reload. }
    \label{fig:meltdown}
\end{figure}

Prior to 2018, the kernel was mapped into the virtual address space of every process, as shown in Figure~\ref{fig:meltdown}. This simplifies system calls and interrupt handling. Since the kernel was in Ring 0, a user function would not be able to directly access the kernel. The Meltdown attack showed how a userspace transient memory load or store operation to a kernel address caused the data to be loaded into the cache memory. This data could then be gleaned using one of the micro-architectural algorithms like Prime+Probe or Flush+Reload (Section~\ref{sec:micro-arch-attacks}).

In the first stage of  Meltdown, the attacker writes code~\cite{Meltdown:2021:code} as shown in Figure~\ref{fig:meltdown} that would perform a load from a kernel address. Specifically {\tt ptr} is made to hold a kernel address. In the ideal case, this should have immediately created an exception because a user instruction is trying to read kernel data. However, modern CPUs are designed in a way that delays the exception, allowing subsequent instructions to be transiently executed. The contents of the kernel space data would thus be loaded into the register {\tt i}, which is then used to load an element from the array into {\tt y}. During this process, {\tt y} is also stored in the cache memory. Notice that the array is indexed based on the kernel data. 
All of these instructions are transiently executed. 
At the time of throwing the exception, the CPU would discard the new values of {\tt i} and {\tt y}, but will not roll back the cache memory. 

In the final stage of Meltdown, either the Flush+Reload or the Prime+Probe can be used to identify the cache set that holds the loaded {\tt array} data, thus revealing information about the kernel data. With the Flush+Reload, for instance, the attacker would first ensure that all {\tt array} elements are flushed from the cache before the transient instructions {\tt M1} and {\tt M2} execute. Post their execution, exactly one element corresponding to {\tt y} would be present in the cache. The cache set that holds {\tt y} can be inferred by measuring the execution time to load each {\tt array} element. The cache set containing {\tt y} would have the shortest load time due to a cache hit. All other elements, by virtue of the initial flush, would result in cache misses.

{\flushleft \bf Spectre.} While the Meltdown attack makes use of an illegal load or store memory operation to induce a transient execution, Spectre makes use of mispredicted branches. Modern microprocessors have a Branch Prediction Unit (BPU) that speculates the direction and the target address of a branch during program execution.
The prediction is done by learning patterns in taken and not-taken branches from the branch history. For example, consider the following code snippet, where {\tt array1\_size} is the size of {\tt array1} and is used to check the bounds of the index {\tt x}. Statements {\tt S2} and {\tt S3} are executed only if {\tt x} is within bounds.  
\begin{verbatim}
    S1. if (x < array1_size){
    S2.    i = array1[x];
    S3.    y = array2[i * 256];
    S4. }
\end{verbatim}
If the snippet is executed repeatedly with legal values of {\tt x}, the BPU would learn the execution pattern  and speculatively execute statements {\tt S2} and {\tt S3}. The results in {\tt i} and {\tt y}, however, would be committed only after the check {\tt x~<~array1\_size} is completed. After a while of such repeated executions, if {\tt x} is made illegal ({\em i.e.} {\tt x~$\tt \ge$~array1\_size}), the BPU would predict incorrectly leading to transiently executed {\tt S1} and {\tt S2}. The two transient memory operations would load data into the cache. The misprediction would ignore the new values computed for  {\tt i} and {\tt y} but not rollback the cache memory.
The final stage of Spectre is similar to Meltdown and uses micro-architectural attack techniques like Evict+Time and Flush+Reload to glean information about  {\tt array1[x]} from the cache memory. For example, if {\tt array1[x]} corresponds to a kernel region, the attack would reveal the contents of the kernel location.

Spectre is one of the most powerful of all transient attacks because it is very difficult to mitigate. Over the years, multiple variants of Spectre have been proposed that exploit the different components of branch speculation in the processor.
The different variants of Spectre attempt to tune different tables in the BPU.  For example,~\cite{Kocher:2019:Spectre,Schwartz:2019:NetSpectre} exploits the Path History Table (PHT), while ~\cite{Bhattacharyya:2019:SMoTherSpectre,Chen:2020:SGXPectre,Kocher:2019:Spectre} exploits the Branch Target Buffer (BTB), and ~\cite{Koruyeh:2018:SpectreReturns,Maisuraze:2012:ret2spec} use the   Return Stack Buffers (RSB).

{\flushleft \bf Foreshadow.} The Meltdown attack breaches the isolation provided by CPU's protection rings there by reading kernel data from a user program. Since 2015, Intel has added another level of protection to its processors. The Intel Security Guard Extension (SGX) is a feature supported by commercial processor variants (deprecated, 11th generation Intel Core onwards~\cite{Intel:2021:Core11Datasheet,Intel:2022:Core12Datasheet}) that provide private regions of memory, called enclaves, for programs. It is ensured that the contents of an enclave, when in RAM, are always in an encrypted form, barring any access to a piece of code outside the enclave, regardless of the privilege levels.
 
The Foreshadow attack makes use of the fact that data in the SGX enclaves are stored in plain form in the L1 cache. This allows transient instructions to compute on the cached secrets. The challenge is to cache secret data in the enclave and use them in transient operations. Given this, the Foreshadow works very similar to Meltdown~\cite{Lipp:2018:Meltdown}. It uses a local buffer that is transiently accessed at indices that depend on secret data stored in the enclave. Now that the entries from the buffer are in the cache, the attacker simply deploys the Flush+Reload technique to establish the secret from the enclave.

In principle, Foreshadow attacks are a variant of the Meltdown attack that uses the same vulnerability, not just to read kernel memory from user space, but to rupture security mechanisms Intel SGX~\cite{Schunter:2016:IntelSGX} that attempt to provide secure enclave protection domains. An improvement on the basic attack is Foreshadow-NG (Next Generation)~\cite{Weisse:2018:ForeshadowNG} which has the potential to read any information that comes to the L1 cache affecting Virtual Machines (VMs), hypervisors (VMM), operating system (OS) kernel memory, and System Management Mode(SMM) memory.

\subsection{Micro-architectural Data Sampling Attacks}
Supporting speculative and out-of-order execution in a microprocessor often requires buffers at several locations in the CPU pipeline that temporarily stores details about in-flight instructions. For example, Reorder Buffers (ROBs) are used to track instructions executed out-of-order and commit their results in the correct order. Other examples are the store buffers, used to track pending stores involved in optimizations.
Micro-architectural Data Sampling (MDS) attacks are able to snoop into these temporary buffers to glean secret data from other applications. 
Unlike Meltdown and Spectre like attacks, MDS attacks are not tied to specific memory addresses, making it almost impossible to mitigate from software. This section summarizes the known MDS attacks.

\begin{figure}[!t]
    \centering
    \includegraphics[width=1.0\linewidth]{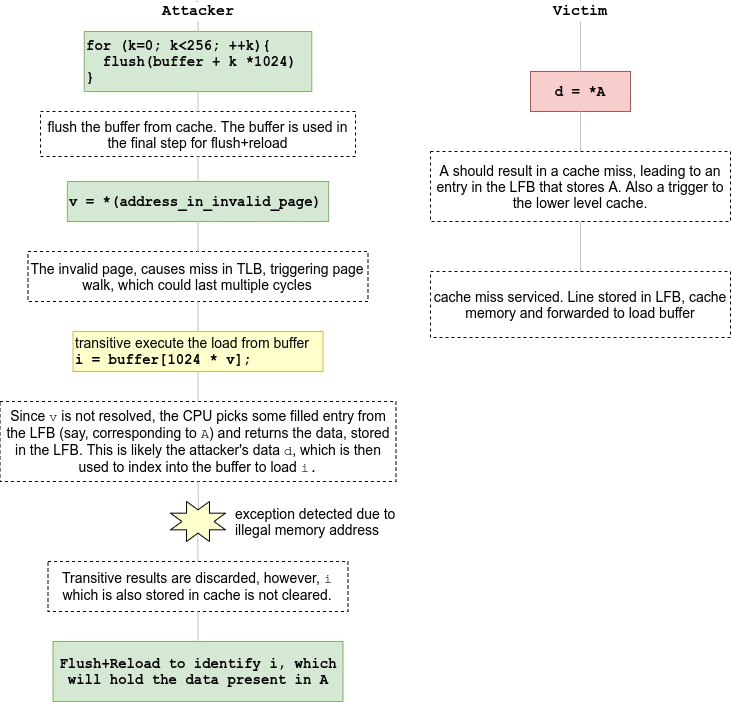}
    \caption{In the RIDL attack, the attacker (in green) snoops into the Line Fill Buffer (LFB) to read the victim's sensitive data present in the address (A).}
    \label{fig:ridl}
\end{figure}

{\flushleft \bf Rogue In-Flight Data Load (RIDL).}
In traditional cache memories, a cache miss would block any further memory requests until the cache miss is serviced. In out-of-order CPUs, addresses corresponding to cache misses are  stored in a Line Fill Buffer (LFB), so that subsequent memory requests can be serviced. This helps create a non-blocking cache. On receiving a memory request that results in a cache miss, an entry in the LFB is created to store the requested address. Subsequently, when the memory block is fetched, it is stored in the LFB entry corresponding to the memory address. The block is also stored in the cache memory and forwarded to the CPU core.
The RIDL attack is able to snoop into the Line Fill Buffer (LFB) to retrieve the data from the stored block. Interestingly, the attack does not depend on the address of the memory request, but only requires a cache miss that makes an entry in the LFB.

RIDL assumes that the attacker and victim share a common L1 cache memory. The steps of the attack are shown in Figure~\ref{fig:ridl}. The attacker first ensures that  {\tt buffer} is flushed from the cache and then triggers the victim to execute a load instruction, say at address {\tt A}. If this victim's load results in a  miss in the L1 cache, a new entry would be created in the LFB which would store the physical address of {\tt A}. 
The attacker, running on a different thread in the same core, issues a load to an address present in a new invalid page. Since this page is new, it would result in a TLB miss and trigger a page table walk. The CPU would eventually detect that the load request is from an invalid page and mark it for an exception. The exception is however thrown much later when  the operation's results are committed in-order. During this time, the memory load operation from {\tt buffer[1024 * v]} would continue transitively using an arbitrary value of {\tt v} picked from an entry in the LFB. The address part of the LFB entry is not matched, therefore, with significant probability, the entry would correspond to the victim's load request at {\tt A}, resulting in {\tt v} holding the value of the victim's data {\tt d}. 
Thus {\tt buffer[1024 * v]} is indexed at a location that is dependent on {\tt d}. The result is stored in {\tt i}, as well as in a cache set.
After the exception is thrown due to the illegal address, the transitive results in {\tt v} and {\tt i} are discarded, however, the cache is not rolled back. Flush+Reload is then used to identify {\tt i}, thus revealing information about the attacker's data.

{\flushleft \bf Zombieload.} 
This attack~\cite{Zombieload:2021:code}, exploits LFB like RIDL, some unknown micro-architectural components and the concept of micro-code assists to mount the attack. Recall that an LFB tracks all load values that are not present in the L1 data cache and needs servicing from higher-level cache hierarchies. Whenever there are complex micro-architectural conditions, such as page faults, they can be handled in one of two ways: (i) The fault can be delegated to a software service routine, or (ii) One can employ \emph{microcode assists}, where the fault is handled through a set of microcode routines, which is faster than delegating to a software. A microcode assist always triggers a pipeline flush resetting the architectural state. However, in-flight instructions still finish execution only to be discarded later. Similarly, the outstanding LFB entries are not discarded. To not incur additional delays in completing the execution of in-flight instructions, the LFB is allowed to load \emph{stale values} for previous load or store instructions, altering the micro-architectural state and potentially allowing the leakage of data. This data can be gleaned by the process of data-sampling explained above. 

Though this attack looks similar to RIDL, the key contrast   of this work is that the above leakage occurs even if the authors systematically ensure using Intel TSX \cite{IntelTSX:2021} that there is no entry filled in Line Buffer during a cache miss. Intel TSX is a set of hardware extensions that enable one program or a program-thread to acquire a lock on certain memory locations in the memory which is prohibited from being updated or used by any other program until released. This enables concurrent programming as the updates in these locations are done atomically by one program or a thread at a time. Within a TSX window, and during certain situations, a miss in the L1-cache never creates a line-buffer entry. However, even without LFB, the leak happens, rather surprisingly at a much higher rate. This suggests that Zombieload is working not only because of LFB but also due to other unknown micro-architectural components, such as FPU register file and store buffer.

{\flushleft \bf Fallout.}
Out-of-order processors hide the latency associated with store operations by using a store buffer. On encountering a store operation, an entry is created in the buffer to hold the virtual address, physical address, and the value to be stored in memory. After the entry is created, subsequent operations in the program can speculatively execute permitting the stores to complete asynchronously.
If one of the subsequent operations is a load, the data from the store buffer is forwarded. This is called {\em store-to-load} forwarding. Such {\em store-to-load} forwarding is possible in two conditions:
\begin{itemize}
    \item  {\bf Condition 1.} If the complete address in the load matches the complete address of an entry in the store buffer, then the value in the entry can be directly used.
    \item {\bf Condition 2.} If the virtual to physical address translation for the load fails, and a few least significant bits match with an entry in the store buffer, then the value in the entry can be speculatively used. 
\end{itemize}
\begin{figure}[!t]
\begin{subfigure}{\textwidth}
  \centering
  \includegraphics[width=.8\linewidth]{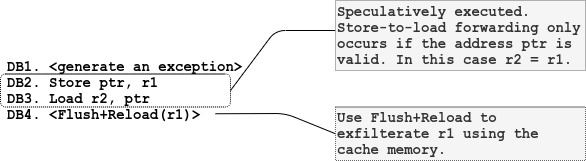}  
  \caption{Data Bounce occurs due to Condition 1.}
  \label{fig:fallout2}
\end{subfigure}
\begin{subfigure}{\textwidth}
  \centering
  \includegraphics[width=.8\linewidth]{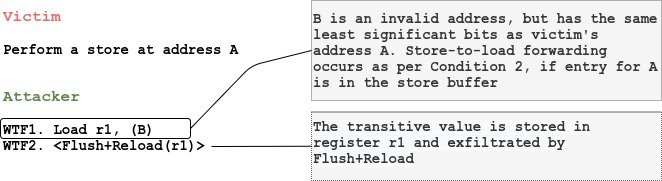}  
  \caption{Write Transitive Forwarding Vulnerability occurs due to Condition 2.}
  \label{fig:fallout1}
\end{subfigure}
\caption{Fallout makes use of Store-to-Load forwarding of data in the store buffer to a speculatively executed load operation. The load operation can be from a different security domain, for example, the kernel. The result of the load is stored in the {\tt r1} register and exfiltrated using Flush+Reload. Flush+Reload is a similar way to previous attacks. The flush is done before the exception-causing instruction, while the reload is done after the transitive execution is discarded.}
\label{fig:fallout}
\end{figure}
In their paper~\cite{Canella:2019:Fallout}, the authors show how both these conditions can lead to transient attacks. The attacks arise from the fact that store-to-load forwarding can happen across security domains. It only requires either of the two conditions to be met. For example, the value in the store buffer entry will be forwarded just by matching address bits in the store buffer entry and the load operation. The store could be from the kernel, while the load is from a user-space program.

The second condition leads to an attack called {\em Data Bounce}, which is used to identify if a virtual address is valid ({\em i.e.} mapped to a physical address). The pseudo-code is shown in Figure~\ref{fig:fallout2}. This attack can be used to break Address Space Layout Randomization (ASLR)~\cite{ASLR:2021,Bhatkar:2003:ASLR,Xu:2003:ASLR}.
The first condition leads to a vulnerability called {\em Write Transient Forwarding (WTF)}. The vulnerability can be used to snoop into stores from another process. Figure~\ref{fig:fallout} provides more details about these attacks.

\begin{figure}[!t]
    \centering
    \includegraphics[width=1.0\linewidth]{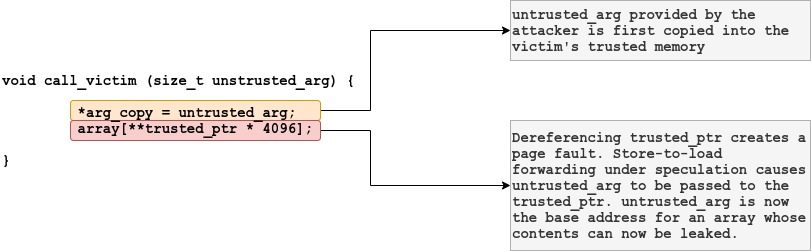}
    \caption{In LVI, the attacker injects a malicious value through load forwarding and uses that to leak sensitive data.}
    \label{fig:lvi}
\end{figure}

{\flushleft \bf Load Value Injection (LVI).} 
In traditional Out-of-order processors, a store to a memory-location followed by a subsequent load instruction to the same memory-location can be  slow as it comprises of two sequential instruction executions involving costly memory accesses. However, a widely used optimization to alleviate this, as explained above in fallout, is to perform \emph{store-to-load} forwarding that will forward the contents of the producing store directly to the consuming load if both the entries are present in the load/store buffer. However, the effective addresses of the load/store instructions are not resolved until later and hence they are speculated instead. Therefore, during speculation, there is a possibility that a wrong store forwards a value to the load. LVI uses this key principle to poison the data that the victim operates on to leak information. This is illustrated using Figure~\ref{fig:lvi}. In this example, the {\tt untrusted\_arg} is sent by the attacker which the victim stores within its buffer space (trusted memory), termed as the poisoning phase. Now, in case there is a page fault caused when dereferencing trusted\_ptr, the {\tt trusted\_ptr} erroneously receives value from the {\tt untrusted\_ptr} due to store-to-load forwarding within the load-store buffers under speculation. This poisoned data is now the index variable for an array whose values can now be leaked through standard cache-based attacks such as Flush + Reload. Generally, the attack contains three phases: (i) Micro-architectural poisoning where the attacker prepares the injection of a poison value by loading that in one of the micro-architectural buffers, (ii) The attacker then provokes the victim into executing instructions that cause a page fault or exception which triggers this store-to-load data poisoning. This can be done, for instance, by evicting a set of victim's virtual memory pages, and (iii) Gadget-based secret transmission, where the attacker finds exploitable code gadgets that can leak data under incorrect transient execution behavior and lead the victim to that code-gadget by carefully poisoning the data.

{\flushleft \bf Crosstalk.} Crosstalk demonstrates that MDS vulnerabilities exist beyond the CPU core through a shared memory buffer, called staging buffer, that is shared across multiple CPU cores. The authors identify several micro-instructions that touch the buffer. These instructions, if executed transiently, can potentially lead to leakage from one CPU core to another. One usecase of Crosstalk is to leak hardware generated random numbers that uses  Intel's Secure Key Technology. The Secure Key technology makes use of an off-core hardware random number generator. The generator is initialized using the {\tt RDSEED} instruction and the random numbers are read using the {\tt RDREAD} instruction. These form the basis of several cryptographic primitives including, Intel's security enclaves.
Executing either of these instructions touches the staging buffer. MDS attacks can be mounted on the buffer by transiently executing {\tt RDRAND} and {\tt RDSEED} thus leaking the seed or the random numbers generated by the hardware random number generator.

\section{Countermeasures}\label{sec:defenses}
Since their discovery, there have been extensive efforts to design and develop countermeasures for transient micro-architectural attacks.
The countermeasures can be broadly classified as prevention-based or detection-based.  Prevention-based solutions attempt to stop the attack by thwarting the execution at one of the three phases (refer Figure~\ref{fig:transient}). Na\" {i}ve  preventive solutions, for instance,  disable speculative execution, thus preventing any transient execution, the first stage of the attack. Another na\" {i}ve preventive solution disables all timers, thus preventing timing channels. This would disable stage 3, {\em i.e.} the transmission of leakage. 
In contrast, detection-based solutions do not disable any feature, rather, they aim to identify patterns in the program execution that can be attributed to an attack. 
While preventive-based solutions have high overheads, detection-based solutions suffer from false positives. 
Over the last few years, there have been multiple detection-based and preventive-based solutions proposed. Table~\ref{tab:defense} provides a list of these solutions. 
This section provides a description and analysis of  some of  these existing solutions.

\begin{table}[]
\centering
\caption{Countermeasures for Transient Micro-Architectural Attacks are classified as either prevention-based or detection-based. While prevention-based techniques aim to either modify or disable some functionality in the software or hardware, detection-based techniques rely on accurately identifying attacks from their run-time characteristics. [HW: Hardware implementation, SW: Software Implementation]
\label{tab:defense}}
\resizebox{\textwidth}{!}{%
\begin{tabular}{|c|c|c|c|c|}
\hline
  \textbf{Stage of} &
  \multirow{2}{*}{\bf Paper} &
  \multirow{2}{*}{\shortstack{\textbf{HW}\\ {\bf or SW?}}} &
  \multirow{2}{*}{\textbf{Threat Model}} &
  \textbf{Reported} \\ 
  \textbf{applicability} &
  \textbf{} &
  \textbf{} &
  \textbf{} &
  \textbf{Overheads} \\ \hline
 \multicolumn{5}{|c|}{\multirow{2}{*}{{\bf --Prevention-based--}}} \\ 
 \multicolumn{5}{|c|}{} \\ \hline
  \multirow{10}{*}{\shortstack{Source \\of Leakage}} &
  NDA~\cite{Weisse:2019:NDA} & \multirow{10}{*}{HW} 
   & 
  Speculative execution attacks & 4-32\%
   \\ \cline{2-2} \cline{4-5} 
                                          & Context~\cite{Schwarz:2020:ConTExT}                 &                                                               & Spectre-like & 0-338\% \\ \cline{2-2} \cline{4-5} 
                                         & InvisiSpec~\cite{Yan:2019:InvisiSpec}              &                                                                & Spectre-like & 5-17\%  \\ \cline{2-2} \cline{4-5} 
                                           & Safespec~\cite{Khasawneh:2019:SafeSpec}                &                                                                & Meltdown, Spectre-like &  3\% \\ \cline{2-2} \cline{4-5} 
                                           & SpectreGuard~\cite{Fustos:2019:SpectrGuard}            &                                                                & Spectre-like & 8-20\% \\ \cline{2-2} \cline{4-5} 
                                           & Specshield ~\cite{Barber:2019:SpecShield}             &                                                                & Speculative execution attacks & 21\% \\ \cline{2-2} \cline{4-5} 
                                           & Spectrum~\cite{Gonzalez:2018:Spectrum}                &                                                                & Spectre-like & 2\% \\ \cline{2-2} \cline{4-5} 
                                           & MuonTrap~\cite{Ainsworth:2020:MuonTrap}                &                                                            & Spectre-like & 4\% \\ \cline{2-2} \cline{4-5}  
                                           & Invisible Speculation ~\cite{Sakalis:2019:invisibleSpeculation}                     &                                                                & Cache and memory side-channels & 11\%  \\ \cline{2-2} \cline{4-5}  
                                          & Reversispec~\cite{Wu:2020:Reversispec}             &                                                                & Speculative load attacks & 8.3\% \\ \cline{1-5} 
  \multirow{8}{*}{\shortstack{Medium\\ of Leakage}}       & Random-fill~\cite{Liu:2014:Random-fill}             & \multirow{8}{*}{HW}                                                               & Contention and reuse based attacks & Negligible  \\ \cline{2-2} \cline{4-5} 
                                           & Newcache~\cite{Liu:2016:Newcache}                &                                                               & Contention and reuse based attacks & Negligible \\ \cline{2-2} \cline{4-5} 
                                           & CEASER~\cite{Qureshi:2018:CEASER}                  &                                                               & Contention-based attacks & 1\% \\ \cline{2-2} \cline{4-5} 
                                           & Encrypted-address cache~\cite{Quresi:2019:Encrypted-address-cache} &                                                            & Contention-bases attacks & 1\%  \\ \cline{2-2} \cline{4-5} 
                                           & Scattercache~\cite{Werner:2019:ScatterCache}            &                                                                & Cache leakage techniques (Section~\ref{sec:micro-arch-attacks}) & 2-4\%  \\ \cline{2-2} \cline{4-5}  
                                           & DAWG~\cite{Kiriansky:2018:DAWG}                    &                                                                & Cache timing attacks & 4-7\% \\ \cline{2-2} \cline{4-5}  
                                           & SecDCP~\cite{Wang:2016:SecDCP}                  &                                                                & Timing side-channels & 12.5\% \\ \cline{2-2} \cline{4-5}  
                                           & MI6~\cite{Bourgeat:2019:MI6}                     &                                                                &  Spectre-like & 16.4\% \\ \cline{1-5} 
  \multirow{4}{*}{\shortstack{Transmission\\ of Leakage}} & Timewarp~\cite{Martin:2012:Timewarp}                & HW                                                               & Timing side-channels & Negligible \\ \cline{2-5} 
                                          & InvarSpec~\cite{Zhao:2020:InvarSpec}        & SW                                                             & Speculative execution attacks & \cite{Yan:2019:InvisiSpec}: 10.9\%  \\ \cline{2-5} 
                                           & oo7~\cite{Wang:2018:oo7}         & SW                                                               & Spectre-like & 5.9\% \\ \cline{2-5} 
                                           & SPECCFI~\cite{Koruyeh:2020:SpecCFI}                     & SW                                                               & Spectre-like & 1.9\% \\ \hline 
 \multicolumn{5}{|c|}{\multirow{2}{*}{{\bf --Detection-based--}}} \\ 
 \multicolumn{5}{|c|}{} \\  \hline 
  \multirow{8}{*}{\shortstack{Transmission\\ of Leakage}} &
  Cyclone~\cite{Harris:2019:D_cyclone} & \multirow{8}{*}{SW}
  & Cache leakage techniques
   & 3.6\%
   \\ \cline{2-2} \cline{4-5} 
                                           & ~\cite{Chiapetta:2016:D_Chiapetta}               & & Cache leakage techniques &  -  \\ \cline{2-2} \cline{4-5} 
                                           & NIGHTs-WATCH~\cite{Mushtaq:2018:Nights_watch}                 &  & Cache leakage techniques & 2\% \\ \cline{2-2} \cline{4-5} 
                                           & WHISPER~\cite{Mushtaq:2020:WHISPER}                 &  & Cache leakage techniques & - \\ \cline{2-2} \cline{4-5} 
                                           & ~\cite{Alam:2021:VictimsCanBeSaviors}                    &  & Cache leakage techniques & - \\ \cline{2-2} \cline{4-5} 
                                          & CloudRadar~\cite{Zhang:2016:CloudRadar}                      &                                                                      & Cross VM attacks & 5\% \\ \cline{2-2} \cline{4-5}  
                                           & CacheShield~\cite{Briongos:2018:CacheShield}                     &                                                                    & Cross VM attacks & - \\ \cline{2-2} 
 \hline
\end{tabular}%
}
\end{table}

\subsection{Prevention-based Countermeasures} 
Figure~\ref{fig:transient} shows the stages of a transient attack. The attacker first identifies a source of leakage, as listed in Table~\ref{tab:tma}. The next step involves the transient movement of data from the source to the medium of leakage. Finally, the attacker uses techniques established in Section~\ref{sec:micro-arch-attacks} to transfer the secret information from the medium. Thwarting any of these sequential stages is sufficient to prevent the attack. Different preventive countermeasures target attacks at different stages of their execution, as described in Table~\ref{tab:defense}. 

Prevention-based countermeasures provide a preemptive solution to these attacks. While the goal of all solutions is to disable the potentially vulnerable behavior of programs, they differ in the attack phase they target. For example,  a preventive solution, called TimeWarp~\cite{Martin:2012:Timewarp} fuzzes the timers in order to prevent attackers from making fine-grained measurements. Such fine-grained measurements are needed to distinguish between micro-architectural events like cache hits and misses.
Without precise time measurements,  the third phase of the attack, namely the flush+reload, would fail. 
While most of these solutions are implemented in the hardware, there are also proposals that work from the software~\cite{Koruyeh:2020:SpecCFI, Wang:2018:oo7, Yan:2019:InvisiSpec}.

{\flushleft \bf Prevention at the source of leakage.}
These countermeasures attempt to thwart attacks at the source of leakage. The most popular approach is to redesign speculative execution in processors to make it leakage-free. A typical solution in this direction divides load instructions into safe and unsafe categories based on the threat model. For example, a load instruction that has committed its results can be considered safe, while a speculative load that is yet to be completed is considered unsafe to prevent Meltdown and  Spectre-like attacks. Countermeasures designed on this philosophy allow the safe loads to alter the global state of the caches. Unsafe loads, however, are not allowed to affect the state of the cache hierarchy~\cite{Ainsworth:2020:MuonTrap,Barber:2019:SpecShield,Gonzalez:2018:Spectrum,Wu:2020:Reversispec,Yan:2019:InvisiSpec}. To implement this, a  buffer is inserted in the processor design that temporally holds the results from speculatively executed instructions until the instruction is completed.

{\flushleft \bf Prevention at the medium of leakage.}
Cache memories store a subset of data in the memory based on temporal and spatial locality. As the cache is several times smaller than the main memory,  multiple addresses map to the same location in the cache resulting in contention. The contention is possible within a process and also across processes. An attacker models this contention to glean information in the cache, using techniques seen in Section~\ref{sec:micro-arch-attacks}. 
Specialized cache memory designs have been proposed for thwarting the attacks by reducing cache contention.
 
In~\cite{percival:05}, Percival suggests eliminating cache contention
by modifying the cache eviction algorithms. The modified
eviction algorithms would minimize the extent to which one thread can evict data from another thread.
In~\cite{page:05}, Page proposed to partition a cache memory that was built of
direct-mapped cache sets that could dynamically be
partitioned into protected regions by the use of specialized cache
management instructions. By tagging memory accesses with partition identifiers, each memory
access is hashed to a dedicated partition. 
While this prevents cache contention from multiple processes,
the cache memory is under-utilized due to rigid partitions. For example, a process may use very
few cache lines of its partition. The unused cache lines are not available to another process.

In~\cite{wang:07}, Wang and Lee provide an improvement on the work by Page~\cite{page:05} using a
construct called {\em partition-locked cache} (PLCache), where
the cache lines of interest are locked in the cache, thereby
creating a private partition. These locked cache lines cannot be evicted by other cache accesses not belonging to the private partition.
In the hardware, each cache line requires additional tags comprising of
a flag to indicate if  the line is locked, and an identifier
to indicate the owner of the cache line. The under-utilization of Page's
partitioned cache still persists because the locked lines cannot be used
by other processes, even after the owner no longer requires them.

In~\cite{domnitser:12}, Domnitser et al. provide a low-cost solution
to prevent attacks based on the fact that the cipher
evicts one or more lines of the spy data from the cache. The solution,
which requires minor modifications of the replacement policies in
cache memories, restricts an application from holding more than
a pre-determined number of lines in each set of a set-associative
cache. With such a cache memory, the spy can never hold all cache
lines in the set, therefore the probability that the cipher evicts
spy data is reduced. By controlling the number of lines that the
spy can hold, a tradeoff between performance and security can be
achieved. Over the years, several other cache partitioning techniques have been suggested ~\cite{Kiriansky:2018:DAWG,Sanchez:2011:cache-partition}  which strengthens the defense while improving usability.

Another well-known modification defense for cache-based attacks makes use of randomization. Wang and Lee propose a {\em random-permutation cache} (RPCache) in~\cite{wang:07},
whereas the name suggests, randomizes the cache interference to make the attack more difficult. The design is based on
the fact that information is leaked only when cache interference is present
between two different processes. RPCache aims at randomizing such
interferences so that no useful information is gleaned. The architecture
requires an additional hardware called the {\em permutation table}, which maps
the set bits in the effective address to obtain new set bits. These
are then used to index the cache set array. Changing the contents
of the permutation table will invalidate the respective lines in the cache.
This causes additional cache misses and randomization in the cache interference. 

An advancement of random cache architectures are designs that encrypt the mapping of addresses to cache sets. CEASER incorporates a block cipher~\cite{Qureshi:2018:CEASER,Quresi:2019:Encrypted-address-cache} for performing the encryption. The encryption key is periodically changed to obtain a different mapping for the cache sets. An important aspect of this design is the encryption algorithm, since it lies in the critical path and influences the time for load and store operations. While traditional ciphers have considerable latencies, ciphers designed specifically for this purpose may not provide sufficiently strong encryption~\cite{Bodduna:20}.

{\flushleft \bf Prevention at the transmission of leakage.}
While there are several techniques to thwart transient attacks by modification in the cache and the execution, existing literature also includes some preventive solution that aims to target the root cause of timing channels, such as fuzzing the timer~\cite{Martin:2012:Timewarp} or increasing the entropy~\cite{Dhavlle:2020:Entropy-Shield} in the timing information. There also solutions that use program analysis~\cite{Wang:2018:oo7,Zhao:2020:InvarSpec} on the program code to identify vulnerable regions and forbid speculative execution in those code sections~\cite{Fustos:2019:SpectrGuard}.

\subsection{Detection-based Countermeasures}

Unlike prevention-based countermeasures, detection-based solutions tend to be reactive in their approach. The detection of any micro-architectural attack involves recognizing some anomalous or malicious pattern of execution. The prevalent technique to classify attacks is to discover features that can provide distinct boundaries between these attacks and benign programs using some statistical method or Machine learning (ML) algorithms. Owing to this, detection-based techniques are more prudent at identifying the transmission of leakage, where the attacker performs distinct operations in the cache to glean the secrets.

A widely popular technique to capture program execution behavior is the use of Hardware Performance Counters (HPCs). These are registers provided by the hardware designer, to monitor certain micro-architectural events in the system. Originally intended for debugging purposes, over the last two decades, HPCs have been shown to profile programs to detect anomalies, malware~\cite{Demme:2013:feasibility_HPC} and specific micro-architectural attacks~\cite{Alam:2021:VictimsCanBeSaviors,Chiapetta:2016:D_Chiapetta,Li:2019:D_HPCsbased,Mushtaq:2018:Nights_watch,Mushtaq:2020:WHISPER,Zhang:2016:CloudRadar}, including those based on transient execution. Such solutions do not detect the anomalies in transient execution, but the step where the attacker gleans the sensitive data.

Another approach to using  HPCs for attack detection is presented by the authors in~\cite{Harris:2019:Cyclone}.  It uses the observation that contention in a resource leaks information only when it is cyclic, meaning domain A interferes with domain B and sequentially domain B interferes with A. Thus the proposal to design a  detection for such cyclic interference patterns using HPCs. While most detection techniques profile the attacker, there are approaches to profile the victim for anomalies~\cite{Briongos:2018:CacheShield}. The end goal of this design is to secure specific domains, rather than a blanket attack detection.

\section{Conclusions}
The last few years have seen several variants of transient micro-architectural attacks. The root cause in all these attacks is the unintended influence of speculatively executed operations with the hardware.
Given the complexity of modern microprocessors, many new variants are likely to be discovered in the future. Next-generation microprocessors should be designed to not just prevent known attacks but should be resilient to future attacks as well. This would require security-aware design methodologies that involve the following.

\begin{itemize}
    \item While there have been several countermeasures proposed, most have been evaluated in an ad-hoc manner. This makes it difficult to quantitatively compare countermeasures and gauge their effectiveness. 
    There is an urgent need to standardize evaluation for security in microprocessors. These standards would provide methodologies to gauge the isolation between software entities. For example, a methodology that can quantify how well the OS is isolated from a userspace program.
    These methodologies could provide toolkits to analyze isolation or a suite of benchmark programs to evaluate the isolation.

    \item Pre and post-Silicon verification of hardware is mainly focused on functional aspects of the design.  Automation tools are designed to minimize area, power, and boot performance. Security vulnerabilities, often fixed in hindsight, have proved expensive. Design automation tools should be augmented to validate for security early in the design phase. This can be a daunting task due to the vast state space of modern microprocessors.
    Artificial Intelligence (AI) is a promising tool that could help design automation for security. Although the use of AI in Electronic Design Automation (EDA) is in its infancy, AI is finding applications to reduce design verification time and achieve more optimized designs. 
    
    \item Proposed preventive countermeasures are designed to stymie specific variants of the attacks. For example, countermeasures for Meltdown are unable to protect against the newer MDS attacks. With multiple attack variants expected in the near future, defense solutions are always catching up with the attacks. 
    
    Detection-based countermeasures, on the other hand, can easily adapt to new attacks. However, most detection countermeasures work from software, and are slow and inaccurate. CPU hardware can be augmented with attack sensors that can detect attacks at runtime with far better accuracy. These sensors should be generic enough to be configured for new attack variants.  
    
    An alternate methodology is to use watchdogs, which monitor processor behavior to detect ongoing attacks. Programmable watchdogs have been proposed in~\cite{Delshadtehrani:2020:PHMon}, and can be extended for micro-architectural attacks.

\end{itemize}

\bibliographystyle{plain}
\bibliography{references}

\end{document}